\documentclass[aps, 12pt]{revtex4}
\usepackage{graphicx,subfigure}
\usepackage{epsfig}
\begin{document}
\title{ A Random Matrix  Approach to Quantum Mechanics }
\author{K.V.S.Shiv Chaitanya $^1$ and B. A.Bambah $^2$}
\email[]{ chaitanya@hyderabad.bits-pilani.ac.in}
\affiliation{$^1$Department of Physics, BITS Pilani, Hyderabad Campus, Jawahar Nagar, \\Shamirpet Mandal,
Hyderabad, India 500 078.}
\affiliation{$^2$School of Physics, University of Hyderabad \\C.R.Rao Road\\Hyderabad Telengana500046.}

\begin{abstract}
We show that the quantum Hamilton Jacobi approach to a class of quantum mechanical bound state problems and the Gaussian orthogonal ensemble of random matrix theory are equivalent. The Berry connection for both problems is identical to their quantum momentum function.The potential that appears in the joint probability distribution function in the random matrix theory is a super potential allowing us to apply it to exceptional polynomials.
\end{abstract}
\maketitle
\section{Introduction}
Since its inception, a wide variety of formulations of non-relativistic quantum mechanics exist \cite{nine}. One can find interesting relationships between the different formalisms. In particular,  two formulations of quantum mechanics, namely, the Quantum Hamilton-Jacobi (QHJ) and the Random Matrix formulation(RMF) bear a striking resemblance. In this paper, we show that solving quantum mechanics using the QHJ formalism is analogous to using random matrix theory.
The probability distribution function of random matrix theory has been used to study a variety of problems in Physics, Chemistry, and  Biology. In physics, in particular, Wigner used random matrices to describe the properties of excited states of atomic nuclei \cite{wig}.  It has also been applied to chiral symmetry breaking in quantum chromodynamics \cite{qcd}, computation of quantum-transport properties of quantum dots \cite{qtqd}, quantum gravity in two dimensions \cite{qg}, the Heisenberg ferromagnet \cite{sas}, quantum chaos \cite{qc}, and in the quantum mechanics of the Calogero-Sutherland model \cite{cs}.

Besides the examples cited above, there are many unusual physical models whose equations resemble those derived from a random matrix ensemble. One such model  is the Stieltjes electrostatic problem \cite{st,st1} which describes $n$ moving unit charges placed
between two fixed charges on a real line, interacting through a logarithmic potential.
  The motion of point vortices in hydrodynamics\cite{aref} $\frac{dz_k}{dt}=\frac{1}{2\pi i}\sum_{j=1}^{n}\frac{\Gamma_j}{z_k-z_j}+W(z_k)$ where the $\Gamma_j$ are the  strengths of point vortices at $z_j$ and $W(z_k)$ is the background flow and the  low-energy spectrum of the Heisenberg ferromagnetic chain \cite{sas}  are more examples. Reference \cite{kvs} shows  the equivalence  between the Stieltjes electrostatic problem and quantum Hamilton
Jacobi (QHJ) formalism and ref \cite{sas}, the connection between the Heisenberg ferromagnetic chain and the Stieltjes electrostatic problem.  Thus, establishing the connection between QHJ formalism and Random Matrix theory is a missing link  and warrants a study.

In this paper, we show such a connection. We use the fact that the probability distribution function of a random matrix is the stationary solution of a Fokker-Planck equation \cite{met} to demonstrate that the probability distribution function of the Gaussian orthogonal ensemble is a wave function of the Schr\"odinger equation.  The probability distribution function of random matrix theory is also shown to be the quantum momentum function.  One of the authors has shown that the quantum momentum function is equivalent to a Berry connection \cite{kvs1}. Hence, probability distribution function is also a Berry connection. 
\subsection{Random Matrix Theory }

 In random matrix theory, the dynamics of the ensemble of a random infinite dimensional Hermitian matrix is described by the probability distribution function (RPDF)
\begin{equation}\label{pdf}
{\cal P}(x_1,\ldots,x_N)\,dx
= c_ne^{-\beta W}
\,\,dx\label{RN}
\end{equation}
where  $dx=dx_1 \ldots dx_N$, $c_n$ is constant of proportionality and the {\it Dyson} index  $\beta=1, 2, 4$  characterizes the  the symmetry class as orthogonal, unitary and symplectic respectively for the random  matrix 
\begin{equation}
W=-\sum_{i=1}^NV(x_i)-\sum_{i<1}^Nln\vert x_i-x_j\vert.
\end{equation}
 
The RPDF can be interpreted as the Boltzmann factor for a classical gas with potential energy $W(x_i,x_j)$ if we identify   $\beta =\frac{1}{T}$ and rescale variables. 
The probability distribution function can be rewritten as 
\begin{equation}
{\cal P}(x_1,\ldots,x_N)\,dx
=c_n e^{-\sum_{i=1}^N\beta V(x_i)}
\prod_{i<j}\vert x_i-x_j\vert^\beta
\,\,dx
\end{equation}
 By manipulating the Vandermonde$\vert x_i-x_j\vert$ determinant, that is, by adding and deleting columns or rows,the probability distribution function can be rewritten 
\begin{equation}\label{pdf1}
{\cal P}(x_1,\ldots,x_N)
=c_N \prod_{i=1}^N w^{\frac{1}{2}} (x_i)
\prod_{i<j}\vert x_i-x_j\vert
\end{equation}
where $w_\beta (x_i)$ is the weight function of classical orthogonal polynomials.
The classical orthogonal polynomials are classified into three different categories depending upon the range of the polynomials. The polynomials in the intervals $(-\infty;\infty)$ with weight function $w(x_i)=e^{-x_i^2}$ are the Hermite polynomials.
In the intervals $[0;\infty)$ with weight function $w=x_i^be^{-x_i}$ are the Laguerre polynomials. In the intervals $[-1;1]$ with weight function $w(x_i)=(1+x_i)^{a}(1-x_i)^b$ are the Jacobi polynomials.For details please refer to \cite{ran2,ran3,ran4,dyson}

The RPDF is the stationary solution of a Fokker-Planck equation \cite{met}.
This can be seen as follows, differentiating (\ref{pdf})w.r.t $x_j$   \begin{equation}
 \frac{\partial P}{\partial x_j} =-\beta P \frac{dW}{dx_j}.
  \end{equation}
Defining \begin{equation}
U(x_j)=-\frac{dW}{dx_j}=\sum_{1\leq j\leq n,j\neq k}\frac{1}{x_k-x_j} + \frac{dV(x_j)}{dx_j},
\end{equation}
we get 
\begin{equation} \sum_{j=1}^N \frac{\partial P}{\partial x_j}-\beta (U(x_j)P)=0.\end{equation} 
Differentiating again w.r.t $x_j$  we get \[\sum_{j=1}^N\frac{\partial }{\partial x_j}\{\frac{1}{\beta} \frac{\partial P}{\partial x_j}-(U(x_j)P)\}=0\] from which it follows that \[\sum_{j=1}^N \frac{1}{\beta} \frac{\partial^2 P}{\partial x_j^2}-\frac{\partial }{\partial x_j}(U(x_j)P)\}=0.\]
Since the Fokker Planck equation is \begin{equation}\label{fp}\frac{\partial P}{\partial t}= \sum_{j=1}^N\{1/\beta \frac{\partial^2 P}{\partial x_j^2}-\frac{\partial }{\partial x_j}(U(x_j)P)\}\end{equation}it follows that $P$  is a stationary solution of the Fokker Planck equation.

\subsection{Quantum Hamilton Jacobi Method }

In this section, we review the Quantum Hamilton Jacobi formalism.
For details see \cite{sree}. 
By defining a function $S$ analogous to the classical characteristic function, related to the wave function $\psi(x,y,z)$ \begin{equation}
\psi(x,y,z) = \exp\left(\frac{iS}{\hbar}\right),       \label{ac}
\end{equation}
and inserting into the Schr\"odinger equation 
\begin{equation}
- \frac{\hbar^2}{2m}\nabla^2\psi(x,y,z)+ V(x,y,z) \psi(x,y,z) = E
  \psi(x,y,z),   \label{sc} 
\end{equation}
 we get
\begin{equation}\label{te}
(\vec{\nabla}S)^2 -i \hbar \vec{\nabla}.(\vec{\nabla}S) = 2m (E
  - V(x,y,z)).   \label{qhj0} 
\end{equation}
The quantum momentum function $p$ is defined as
\begin{equation}
\vec{p} = \vec{\nabla} S. \label{mp}
\end{equation} 
In terms  $\vec{p} $ , the QHJ equation is  
\begin{equation}
(\vec{p})^2 - i \hbar \vec{\nabla}.\vec{p} = 2m (E - V(x,y,z)).
\label{bhy} 
\end{equation} 
\begin{equation}
\vec{p} = -i \hbar \vec{\nabla} ln \psi(x,y,z) \label{lg}
\end{equation}

In one dimension \begin{equation}
p = -i \hbar \frac{d}{dx}ln \psi(x). \label{lg1}
\end{equation}
and 
\begin{equation}
p^2 - i \hbar \frac{dp}{dx} = 2m (E - V(x)),       \label{qhj1}
\end{equation}
which is known as the Riccati equation. 
Including   time dependence in (\ref{te})
\begin{equation}
i  \frac{\partial }{\partial t}S (x,t)=(\frac{\partial S}{\partial x})^2 - i  \frac{d^2S}{dx^2} - 2m (E - V(x)),   \label{qhj14}   
\end{equation}
and differentiating with respect $x$, we get
 \begin{eqnarray}
i S_{xt}(x,t)=-i\frac{\partial^3  S}{\partial x^3}-2\frac{\partial S}{\partial x}(\frac{\partial^2 S}{\partial x^2})+\frac{\partial V}{\partial x}.
\end{eqnarray} 

In terms of  $p(x,t)=\frac{\partial S}{\partial x}$
 \begin{eqnarray} \label{bh}
i \frac{\partial }{\partial t}p (x,t)=-i\frac{\partial^2 p}{\partial x^2}-2 p(\frac{\partial p}{\partial x})+\frac{\partial V}{\partial x},\label{bj}
\end{eqnarray} 
 which is the the  Burger-Hopf equation.
It has been  shown by Leacock and Padgett \cite{qhj} that the action angle variables 
give rise to exact quantization condition
\begin{equation}
J(E) \equiv  \frac{1}{2\pi} \oint_C{pdx} = n\hbar.       \label{act}
\end{equation}
In the next section we show that  the solutions of  Burger-Hopf equation through the Cole-Hopf transform  map  to  the Fokker Planck Equation, the Random Matrix theory and Supersymmettric quantum mechanics.

\section{Connection between RMT, and SUSY QM QHJ}
We have shown that probability distribution of random matrix is a stationary solution to the Fokker-Planck equation. Here we would like to map the problem to the quantum Hamilton Jacobi problem. In quantum mechanics, in general, we solve an eigenvalue equation for a self-adjoint operator. The time dependent form of the Fokker-Planck equation satisfied by the Probability Distribution function( \ref{fp}),can be written as 
\begin{equation}\label{fp2}
\frac{\partial P}{\partial t}=\sum_{j=1}^N\{1/\beta \frac{\partial^2 P}{\partial x_j^2}-\frac{\partial }{\partial x_j}(U(x_j)P)\}
\end{equation}
where\[U(x_j)=-\frac{dW}{dx_j}=\sum_{1\leq j\leq n,j\neq k}\frac{1}{x_k-x_j} + \frac{dV(x_j)}{dx_j}.\]
If we define the action $S$  in terms of the conditional probability P as 
 \begin{equation}\label{fp1}
P(x_j,t)=N(t) e^{-\beta S(x_j,t)+\frac{\beta}{2} W(x_j)}.
\end{equation} 
The equation satisfied by ({\ref{fp1}) is 
\begin{eqnarray}
\frac{\partial P}{\partial t}&=&\left(\frac{\dot{N}}{N}-\beta\frac{\partial S}{\partial t} \right) P \\
\frac{\partial P}{\partial x_j}&=&-\beta P \frac{\partial S}{\partial x_j}+\frac{\beta}{2} P \frac{\partial W}{\partial x_j}\\
\frac{\partial^2 P}{\partial x_j^2}&=&-\beta P \frac{\partial^2 S}{\partial^2 x_j}
+\frac{\beta}{2} P \frac{\partial^2 W}{\partial^2 x_j}+\beta^2 P(\frac{\partial S}{\partial x_j})^2
+\frac{\beta^2}{4} P(\frac{\partial W}{\partial x_j})^2-\beta^2 P(\frac{\partial S}{\partial x_j})(\frac{\partial W}{\partial x_j})
\end{eqnarray} 

Thus the FP equation  (\ref{fp2}) becomes
\begin{eqnarray}
\left((\frac{\dot{N}}{N}-\beta\frac{\partial S}{\partial t} \right) P&=&\sum_j(-\beta P \frac{\partial^2 S}{\partial^2 x_j}
+\frac{\beta}{2} P \frac{\partial^2 W}{\partial^2 x_j}+\beta^2 P(\frac{\partial S}{\partial x_j})^2
+\frac{\beta^2}{4} P(\frac{\partial W}{\partial x_j})^2\nonumber\\&&-\beta^2 P(\frac{\partial S}{\partial x_j})(\frac{\partial W}{\partial x_j})-\beta\frac{\partial U}{\partial x_j} P+\beta^2 UP\frac{\partial S}{\partial x_j}-\frac{\beta^2}{2} UP\frac{\partial W}{\partial x_j})
\end{eqnarray}

Since $U=\frac{\partial W}{\partial x_j}$ 
\begin{eqnarray}
\left((\frac{\dot{N}}{N}-\beta\frac{\partial S}{\partial t} \right) P&=&\sum_j(-\beta  \frac{\partial^2 S}{\partial^2 x_j}
+\frac{\beta}{2}  \frac{\partial^2 W}{\partial^2 x_j}+\beta^2 (\frac{\partial S}{\partial x_j})^2
+\frac{\beta^2}{4} (\frac{\partial W}{\partial x_j})^2\nonumber\\&&-\beta\frac{\partial^2 W}{\partial x_j^2}  -\frac{\beta^2}{2} \frac{\partial W}{\partial x_j})P.
\end{eqnarray}
 For constant $N$ we are led to
\begin{eqnarray}
\frac{\partial S}{\partial t}P &=&\sum_j( \frac{\partial^2 S}{\partial^2 x_j}
+\frac{1}{2}  \frac{\partial^2 W}{\partial^2 x_j}-\beta(\frac{\partial S}{\partial x_j})^2
-\frac{\beta}{4} (\frac{\partial W}{\partial x_j})^2)P
\end{eqnarray}

Defining the momentum function to be $p(x)=\frac{\partial S}{ \partial x}$ then the equation is precisely eqn.(\ref{bh}) the Burger-Hopf form of the quantum Hamilton Jacobi equation.

Now we go from QHJ to RMT. The Burger-Hopf equation admits a solution  with  the following pole structure  as is shown in \cite{cho}
\begin{equation}
p(x)=\sum_{k=1}^n\frac{-i\hbar}{x-x_k}+Q(x).\label{uf0}
\end{equation}
By introducing the polynomial
\begin{eqnarray}
f(x)=(x-x_1)(x-x_2)\cdots (x-x_n),\label{poly}
\end{eqnarray}
 the quantum momentum function ( $\hbar=1$) is
\begin{equation}
p=- i\frac{f'(x)}{f(x)}+Q(x)\label{uf4}.
\end{equation}
Substituting in (\ref{qhj1}) one gets
\begin{eqnarray}
-f''(x) + 2iQ(x)f'(x)+[Q^2(x)-iQ'(x)-E+V(x)]f(x)=0.\label{dif11}
\end{eqnarray} 
The search for the polynomial solutions to the equation (\ref{dif11}) leads to quantization. This is equivalent to demanding $ [Q^2(x)-iQ'(x)-E + V(x)]$ to be constant. This will only be possible  if 
$Q(x)=i\mathcal{W}(x)$.  In QHJ, it turns out that the $Q(x)$ is related to  the super potential.

In supersymmetric quantum mechanics , the superpotential $\mathcal{W}(x)$ is defined in terms of the intertwining operators $ \hat{A}$ and $\hat{A}^{\dagger}$  as
\begin{equation}
  \hat{A} = \frac{d}{dx} + \mathcal{W}(x), \qquad \hat{A}^{\dagger} = - \frac{d}{dx} + \mathcal{W}(x), 
\label{eq:A}
\end{equation}
This allows one to define a pair of factorized Hamiltonians $H^{\pm}$ as
\begin{eqnarray}
   H^{+} &=&     \hat{A}^{\dagger} \hat{A}     = - \frac{d^2}{dx^2} + \mathcal{V}^{+}(x) - E, \label{vp}\\
  H^{-} &=&     \hat{A}  {\hat A}^{\dagger}     = - \frac{d^2}{dx^2} + \mathcal{V}^{-}(x) - E, \label{vm}
\end{eqnarray}
where $E$ is the factorization energy. 
The partner potentials $\mathcal{V}^{\pm}(x)$ are related to $\mathcal{W}(x)$ by 
\begin{equation}\label{gh}
 \mathcal{V}^{\pm}(x) = \mathcal{W}^2(x) \mp \mathcal{W}'(x) + E.
\end{equation}
Thus the identification $Q(x)=i\mathcal{W}(x)$ in equation (\ref{dif11})establishes the connection with the superpotential. We refer the reader to \cite{kharebook} for more details on SUSY QM. We also note the similarity between the super potential  and the Potential of the Shr\"odinger wave equation satisfied by the Random matrix probability function.
 
The Fokker Planck operator can be mapped into the Shr\"{o}dinger equation in the following way. The Fokker Planck operator can be written as \[ L=\sum_{j=1}^N\frac{\partial }{\partial x_j}\{\frac{1}{\beta} \frac{\partial }{\partial x_j}+\frac{\partial W}{\partial x_j}\} \]  so that, \[ L=\sum_{j=1}^N\frac{\partial e^{-\beta W} }{\partial x_j}\{\frac{1}{\beta} \frac{\partial e^{\beta W} }{\partial x_j}\} \] 
 $e^{\beta W/2}Le^{-\beta W/2}$ is a Hermitian operator which is
\[e^{\beta W/2}Le^{-\beta W/2}=\sum_{j=1}^N\frac{1}{\beta} \frac{\partial^2 }{\partial x_j^2}-\frac{1}{2}\frac{\partial^2 W(x_j)}{\partial x_j^2}-\frac{\beta}{2}(\frac{\partial W(x_j)}{\partial x_j})^2.\]
Writing \[P=e^{-\beta W/2}\psi\]  and substituting into the Fokker Plank Equation leads us to $\psi$ being a solution to the Shr\"{o}dinger equation with potential given by $V=\frac{1}{2}\frac{\partial^2 W(x_j)}{\partial x_j^2}+\frac{\beta}{2}(\frac{\partial W(x_j)}{\partial x_j})^2=$ .
For the Time dependent Fokker Planck Equation we substitute$ P=e^{-E t}e^{-\beta W/2}\psi$ then the Fokker Planck Equation becomes \[i\frac{\partial \psi}{\partial t}=-V\psi+\frac{1}{2 \beta}\frac{\partial^2 \psi}{\partial x_j^2}\]  which is precisely the Schr\"{o}dinger equation for an appropriately chosen $\beta$.

The complex pole expansion of the solution of  Burger's equation (\ref{uf0}), for $Q(x)=0$,  is given by
\begin{equation}
p(x)=\sum_{k=1}^n\frac{-i}{x-x_k}.\label{ufo}
\end{equation}
It is proved in ref\cite{cho} that this Burger's equation is satisfied if and only if the complex poles evolve according to the system of $n$ linear equations \cite{aref} given by
\begin{equation}
\sum_{j=1}^n \frac{d x_j}{dt}=\sum_{1\leq j\leq n,j\neq k}^n\frac{i}{x_k-x_j},\label{ufie}
\end{equation}
 in each equation $j=k$ term is not present. For the stationary case $\frac{d x_j}{dt}=0$ 
 \begin{equation}
\sum_{1\leq j\leq n,j\neq k}^n\frac{i }{x_k-x_j}=0,\label{ufiek}
\end{equation}
Including the potential \begin{equation}
\sum_{1\leq j\leq n,j\neq k}^n\frac{-i}{x_k-x_j}+\sum_{j=1}^n Q(x_j)=0.\label{ufiu}
\end{equation}
We identify equation (\ref{ufiu}) with the quantum momentum function
\begin{equation}
p(x_j)=\sum_{1\leq j\leq n,j\neq k}^n\frac{-i}{x_k-x_j}+\sum_{j=1}^n Q(x_j).\label{ufi}
\end{equation}
 and $p(x_j)=0$ satisfies the system of $n$ linear equations. The potential $Q(x_j)$ appearing in quantum momentum function (\ref{ufi}) is super potential. This fact is later used to derive the probability distribution function of random matrix to be the wave function of the Schr\"odinger equation.

In order to compare the RMT and QHJ, we define the probability distribution function in RMT as the wave function 
\begin{equation}
\psi(x_k)={\cal P}(x_1,\ldots,x_N)=C_n exp^{-\beta W}\label{rwf}
\end{equation}
where 
\begin{equation}
W=V(x_k)-\frac{1}{2}\sum_{1\leq j\leq n,j\neq k}ln(\vert x_k-x_j\vert)
\end{equation} 
Substituting (\ref{rwf}) in (\ref{lg1}) the quantum momentum function is given by
\begin{eqnarray}\label{uf11}
p(x_k)=\sum_{1\leq j\leq n,j\neq k}\frac{\beta}{x_k-x_j}-S(x_k)\;\; k=1,2\cdots n
\end{eqnarray}
where $S(x_k)=\beta\frac{dV(x_l)}{dx_k}$. By comparing equation (\ref{uf11}) and (\ref{uf0}) one gets $\beta=-i \hbar$ and 
$\mathcal{W}(x)=\frac{1}{\beta}S(x)=iQ(x)$ ($\hbar=1$). It is therefore clear that the potential used in the random matrix theory is the super potential $\mathcal{W}(x)$.
 The equation (\ref{uf11}) is the system of $n$ linear equation, which is identical to the Steiljes electrostatic model as shown in Metha's Book \cite{met}. 

For Gaussian Ensembles 
 \begin{equation}\label{pdf0}
 P(x_1,x_2\cdots x_n)=\frac{1}{C_{\beta}}e^{-\beta(\sum_{j=1}^n x_j^2+\sum_{j,k}ln(x_j-x_k))}, \end {equation}
 which is the Partition function for a Coulomb Log-gas , as was pointed out by Dyson \cite{dyson}.
 The Schr\"odinger equation for the Harmonic oscillator potential admits the Dyson gas probability distribution as the wave function. Consider the Harmonic oscillator potential $V(x_j)=\frac{1}{2} x_j^2$, such that, the superpotential is $Q(x_j)= x_j$. The residue of fixed poles is 
$\pm i$, we pick $+i$ as it corresponds to physical states. Using the relationship between the wave function and quantum momentum function 
 \begin{equation}
\frac{d}{dx_j}ln \psi(x_j)=\sum_{1\leq j\leq n,j\neq k}^n\frac{1}{x_k-x_j}-\sum_{j=1}^nx_j,\label{ufip}
\end{equation}
 Integrating the equation (\ref{ufip}) gives the wave function
\begin{equation}
\psi(x_j)=\prod_{N \geq k> j \geq 1}\left( x_k-x_j \right) \prod^N_{j=1}\exp\left( -  \frac{1}{2} x_i^2 \right).\label{ufip11}
\end{equation}
Therefore, we have shown that the Schr\"odinger equation admits the probability distribution function of Gaussian orthogonal ensemble(\ref{pdf}) for the Dyson log gas model in random matrix theory \cite{met}. For a general potential the wave function is given by
\begin{equation}
\psi(x_j)=\prod_{N \geq k> j \geq 1}\left( x_k-x_j \right) \prod^N_{j=1}\exp\left( - i\int Q(x_i) dx_j \right), \label{ufip1p1}
\end{equation}
here $ Q(x_j)$ is super potential.
The quantum momentum function is defined if the moving poles are simple poles. It should be clear from the probability distribution function in equation (\ref{pdf0}), the moving poles will be simple poles for the Gaussian orthogonal ensemble. 
By introducing the polynomial $f(x)$  in the quantum momentum function ( $\hbar=1$) to get the momentum function
$p(x)=- i\frac{f'(x)}{f(x)}+Q(x)$ 
and substituting  into the  Riccati equation one gets
the following differential equation  as
\begin{eqnarray}
-f''(x) + 2iQ(\lambda)f'(\lambda)+[Q^2(\lambda)-iQ'(\lambda)-E+V(\lambda)]f(\lambda)=0.\label{fgy}
\end{eqnarray} 
which is exactly same as quantum Hamilton Jacobi equation (\ref{dif11}). Thus our objective of establishing a connection between Random Matrix theory, Quantum Hamilton Jacobi Method and Supersymmetric quantum mechanics is achieved.

\section{ Examples}
As an illustration of the equivalence described, we show that a class of  bound state problems of quantum mechanics can be mapped to random matrix ensemble . 

First, consider the Coulomb potential in natural units
\begin{equation}
V_{coul}(x)=\frac{l(l+1)}{r_j^2}-\frac{1}{r_j}+E.
\end{equation}
The super-potential for the Coulomb potential is 
\begin{equation}
\mathcal{W}_{coul}(x)=\frac{1}{2}-\frac{(l+1)}{r_j}.
\end{equation}
The Hamiltonian in random matrix theory  is 
\begin{equation}
W=-\sum_{1\leq j\leq n,j\neq k}ln\vert r_k-r_j\vert-V(r_j),\label{w}
\end{equation}
 which has a  minimum when
\begin{equation}
\sum_{1\leq j\leq n,j\neq k}\frac{1}{ r_k-r_j}-\frac{\partial V(r_j)}{\partial r_j}=0.\label{rmu}
\end{equation}

Using the identity \begin{eqnarray}
\sum_{k=1}\frac{1}{r-r_k}=\frac{f'(r)}{f(r)}-\frac{1}{r-r_j}=\frac{(r-r_j)f'(r)-f(r}{(r-r_j)f(x)},\label{dig}
\end{eqnarray}
and  taking the limit $r\rightarrow r_j$ and using l'Hospital rule one gets
\begin{eqnarray}
\left[\frac{f'(r)}{f(r)}-\frac{1}{r-r_j}\right]&=&\lim_{r\rightarrow r_j}\frac{(r-r_j)f'(r)-f(r)}{(r-r_j)f(r)}
\nonumber \\&=& \frac{f''(r_j)}{2f'(r_j)}.\label{lhp}
\end{eqnarray}
Substituting  equation (\ref{lhp}) in the equation (\ref{rmu}) 
\begin{equation}
\frac{f''(r_j)}{2f'(r_j)}-(\frac{1}{2}-\frac{(l+1)}{r_j}).=0.
\end{equation}
then
\begin{equation}
r_jf''(r_j)+(2(l+1)-r_j)f'(r_j)=0.
\end{equation}

This equation has stable equilibrium at the zeros of the Laguerre orthogonal polynomial and the general differential equation is given by
\begin{equation}
r_jf''(r_j)+(2(l+1)-r_j)f'(r_j)+\lambda f(r_j)=0.
\end{equation}
Then the solution for $f(r)=L^(2l+1)_\lambda(r)$, $\lambda=n$ being an integer.
The corresponding wave function is $\psi(r)= r^{l+1}exp[-\frac{1}{2}r] L^{2l+1}_n(r)$. Hence, $\psi(r)$ is written in the in the form of equation (\ref{pdf1}).

The 3d Harmonic oscillator.
\begin{equation}
V_{osc}(x)=\frac{l(l+1)}{x^2}+\frac{x^2}{4}-E,
\end{equation}
can be mapped by the transformation 
$x^2=r$ to the Coulomb problem
\begin{equation}
V_{coul}(r)=\frac{1}{r}V_{osc}(r)=\frac{l(l+1)}{r^2}+\frac{1}{4}-\frac{E}{r}.
\end{equation}
and all the arguments for the Coulomb problem can be used to show that the wave function is of the type (\ref{pdf1}).

Similarly, the Morse potential 
\begin{equation}
V_{mor}(r)=A^2+B^2exp(-2\alpha x)-2B(A+\alpha/2)exp(-\alpha x)
\end{equation}
can be mapped by the  transformation
$r=exp(-\alpha x)$ to the Coulomb potential
\begin{equation}
V_{mor}(r)=A^2+\frac{B^2}{r^2}-\frac{2B(A+\alpha/2)}{r}
\end{equation}

Similarly, the Scarf Potential
\begin{eqnarray}
V(x)&=&-A^2+(A^2+B^2-A\alpha)sec^2(\alpha x)-B(2A-\alpha)tan(\alpha x)sec(\alpha x)
\end{eqnarray}
is transformed  by $sin(\alpha x)=\frac{1}{r}$, to
\begin{eqnarray}
\frac{r^2-1}{r}V_{scarf}(x)&=&\frac{A^2}{r^2}+(B^2-A\alpha)-\frac{B(2A-\alpha)}{r}
\end{eqnarray}
 which is related to the Coulomb potential by $V_{coul}(r)=\frac{r^2-1}{r}V_{Scarf}(r)$.
 
To see how the RPDF can be related to exceptional polynomials, we note that
the quantum Hamiltonian-Jacobi formalism for the deformed oscillator in ref \cite{pkp} gives the momentum function as
\begin{eqnarray}\label{hj}
p=\sum_{k=1}^{n}\frac{1}{x-x_k}-x-\frac{1}{x-x_1}+\frac{g+l}{x}.
\end{eqnarray}
The solutions of the corresponding differential Burger -Hopf equation  are the exceptional Laguerre polynomials 
\begin{eqnarray}
\psi=\frac{x^{g+l}e^{-\frac{1}{2}x^2}}{L^{g+l - \frac{3}{2}}_l(−x^2)}\hat{L}^n_l(x^2,g).
\end{eqnarray}
By following above procedure the joint probability distribution function for exceptional Laguerre polynomials is given by
\begin{eqnarray}\label{pdf11}
{\cal P}(x_1,\ldots,x_N)\,dx
=C_n \frac{x^{2(g+l)}e^{- x^2}}{L^{2(g+l)-3}_l(−x^2)}
\prod_{i<j}\vert x_i-x_j\vert^2
\,\,dx.
\end{eqnarray} 
where $\hat{L}^n_l(x^2,g)$ are exceptional Laguerre polynomials. 
 \section{Quantum Momentum Function and Berry Phase}
  It has been shown by one of the authors that the equation (\ref{ufi}) is the Berry connection in  the context of studying the Laughlin wave function\cite{kvs1}. For completeness, we prove that the quantum momentum function is the Berry connection. 
  If 
 \begin{equation}\label{brt}
 A(r)=i\langle\psi\vert\nabla\vert\psi\rangle,
 \end{equation}
 where $\vert\psi\rangle$ is the wave function , then  Berry phase is defined as
 \begin{equation}
 \gamma=-i\oint_C  A(r) dr.
 \end{equation}
Substituting equation (\ref{ac}) in equation (\ref{brt}),   gives 
$A(r)=\nabla S$, where $S$ is considered in three dimensions. Since, $A(r)$ is a like a vector potential and is equal to the gradient of the action $S$, hence, $\nabla\times \nabla S =0 $. If $S$ has singularities, then
$\nabla\times \nabla S\neq 0 $. The relation between the momentum function and action is given by the equation (\ref{mp}), in three dimensions $p=\nabla S$, identical with the Berry Phase.
If we are working in one dimension we calculate the Berry phase in one dimension. 
To calculate Berry phase, we use the Stokes theorem given by 
\begin{equation}
 \gamma \equiv  \frac{1}{2\pi} \oint_C{\nabla\times pdS} =\frac{1}{2\pi} \oint_C{pdx} = n\hbar.       \label{act1}
\end{equation}
 Once these singularities are quantized the connection is exact. This shows us that the quantization is related to the topology of the underlying  system.  Hence, the quantum momentum function defined in the equation (\ref{ufi}) and  the minimum of the random matrix defined in equation (\ref{uf11}), when they are identical gives the same quantization condition.
 
Therefore, the quantum mechanics and the Gaussian orthogonal ensemble of random matrix theory are topologically equivalent as the Berry connection for both the class of problems is identical to the quantum momentum function.
 
\section{Conclusion}
In this paper, we gave a random matrix theory approach to the quantum mechanics using the quantum 
Hamilton-Jacobi formalism. We have shown that the bound state problems in quantum mechanics are analogous to solving Gaussian unitary ensemble of random matrix theory. We have also shown that the potential appears in the joint probability distribution function in the random matrix theory as a super potential. Using this approach we have extended the random matrix theory to the newly discovered exceptional polynomials.
We have also shown that the quantum mechanics bound state problems and the Gaussian orthogonal ensemble of random matrix theory are topologically equivalent as the Berry connection for both the class of problems is identical to the quantum momentum function.
\section*{Acknowledgments}
KVSSC acknowledges the Department of Science and Technology, Govt of India (fast-track scheme (D. O. No: SR/FTP/PS-139/2012)) for financial support.

\end{document}